\documentclass[twocolumn,prd,showpacs,amsmath,amssymb,aps]{revtex4}

\usepackage{graphicx,latexsym,amsmath,amssymb}
\usepackage{dcolumn}

\newcommand{\beq}{\begin{equation}}
\newcommand{\eeq}{\end{equation}}
\newcommand{\be}{\begin{eqnarray}}
\newcommand{\ee}{\end{eqnarray}}

\newcommand{\rar}{\rightarrow}
\newcommand{\Rar}{\Rightarrow}

\def\dd{ \,\mathrm{d} }

\def\+{\dagger}
\def\la{\langle}
\def\ra{\rangle}
\def\<{\langle}
\def\>{\rangle}

\newcommand{\exclude}[1]{}
\newcommand{\Lqcd}{\Lambda_{\mathrm{QCD}}}
\newcommand{\Tr}{\mathrm{Tr}}

\begin{document}

\title{Cosmological constant, violation of cosmological isotropy and CMB.}

\author{Federico R. Urban and Ariel R. Zhitnitsky}

\affiliation{Department of Physics \& Astronomy, University of British Columbia, Vancouver, B.C. V6T 1Z1, Canada}

\date{\today}

\begin{abstract}
We suggest that the solution to the cosmological vacuum energy puzzle does not require any new field beyond the standard model, but rather can be explained as a result of the interaction of the infrared sector of the effective theory of gravity with standard model fields.  The cosmological constant in this framework can be presented in terms of QCD parameters and the Hubble constant $H$ as follows, $\epsilon_{vac}  \simeq H \cdot  m_q\la\bar{q}q\ra  /m_{\eta'} \simeq  (4.3\cdot 10^{-3} \text{eV})^4$, which is amazingly close to the observed value today.  In this work we explain how this proposal can be tested by analyzing CMB data.  In particular, knowing the value of the observed cosmological constant fixes univocally the smallest size of the spatially flat, constant time 3d hypersurface which, for instance in the case of an effective 1-torus, is predicted to be around 74 Gpc. We also comment on another important prediction of this framework which  is a violation of cosmological isotropy.  Such anisotropy is indeed apparently observed by WMAP, and will be confirmed (or ruled out) by future PLANCK data.
\end{abstract}

\maketitle

\section{Prelude}

It has been suggested recently~\cite{our4d} that the solution of cosmological vacuum energy puzzle may not require any new field beyond the standard model.  The idea is based on the philosophy that gravitation can not be a truly fundamental interaction, but rather it must be considered as a low energy effective quantum field theory (QFT)~\cite{Thomas:2009uh}.  In such a case, the corresponding gravitons should be treated as quasiparticles which do not feel all the microscopic degrees of freedom, but rather are sensitive to the ``relevant excitations'' only. In this framework it is quite natural to define the ``renormalised cosmological constant'' to be zero in Minkowski vacuum   wherein the Einstein equations are automatically satisfied as the Ricci tensor identically vanishes in flat space. Thus, the energy momentum tensor   in combination with this ``bare cosmological constant'' must also vanish at this specific ``point of normalisation'' to satisfy the Einstein equations.  From this definition it is quite obvious that the ``renormalised energy density'' must be proportional to the deviation from Minkowski space time geometry.  With this definition the effective QFT of gravity has a predictive power.   In particular, it predicts the behaviour of the system in any non-trivial geometry of the space time. 

The first application of this proposal was the computation of the cosmological constant in a spacetime with non-trivial topological structure~\cite{our4d}.  It was shown that the cosmological constant does not vanish if our universe can be represented by a large but finite manifold with typical size $L\simeq (cH)^{-1}$ where $c$ is a coefficient of order 1, and where $H$ is the Hubble constant.  The cosmological vacuum energy density $\rho_{\Lambda} $ in this framework is expressed in terms of QCD parameters for $N_f=2 $ light flavours as follows:
\be
\label{rho}
\rho_{\Lambda} \simeq 
c\cdot \frac{2H N_f  |m_q\la\bar{q}q\ra  |}{m_{\eta'}} \simeq c\cdot (4.3\cdot 10^{-3} \text{eV})^4 \, .
\ee
This estimate should be compared with the observational value $ \rho_{\Lambda} = (2.3\cdot 10^{-3} \text{eV})^4$.  The deviation of the cosmological constant from zero is entirely due to the large but finite size $L$ of the manifold.  The non-vanishing result~(\ref{rho}) can be understood as a Casimir type of vacuum energy when the boundary conditions and topology play a crucial r\^ole.

It is also interesting to note that a somewhat similar estimate was given in 1967 by Zeldovich~\cite{Zeldovich:1967gd} who argued that  $\rho_\Lambda \simeq {m_p^6}/{M_P^2}$: this is numerically of the same order of magnitude, and has the same form of~(\ref{rho}) if one replaces $\Lqcd \rar m_p$ ($\Lqcd$ was not known at that time) and $H \rar \Lqcd^3/ M_P^2$.  Since then, the form $H\Lqcd^3$ has risen its head several times, see for instance~\cite{Bjorken:2001pe,Schutzhold:2002pr,Bjorken:2004an,Klinkhamer:2007pe,Klinkhamer:2008ns,Klinkhamer:2009nn}.  Despite the apparent similarity between the form $H\Lqcd^3$ and eq.~(\ref{rho})--but see~(\ref{vacuum}) for the precise result--it is important to notice that, first of all, in our case it is the \emph{inverse linear size of the embedding compact manifold} which actually appears in the exact result~(\ref{vacuum}) below, not the Hubble parameter, as in all the aforementioned papers.  This is because we work on a compact manifold of characteristic size $L$, and we use the replacement $H\simeq L^{-1}$ only for estimation purposes (e.g., we want to make sure that our manifold does not turn out to be much smaller than the last scattering surface). Therefore, our result is of fundamentally different origin.  Finally, let us stress that equation~(\ref{rho}), or to be more precise, eq.~(\ref{vacuum}), is not an estimate based on (although very well motivated) physical arguments, but a result of a precise calculation, which can be explicitely worked out completely analytically in a simplified 2d model defined on a finite manifold such as a 2-torus~\cite{toy}.

In this paper we argue that the estimate~(\ref{rho}), and in turn the basis upon which it is founded, can be confronted against observations by analysing cosmic microwave background (CMB) data, which is already quite sensitive to the relevant scales.
  
The next section is devoted to a short introduction on finite, topologically non-trivial manifolds which are relevant in our discussions.  We also give a brief review on how topological structures can be looked for in the CMB.  Section III is a short overview of our work~\cite{our4d}, in which eq.~(\ref{rho}) was originally derived.  Finally, in section IV we describe in detail the specific CMB signatures of our proposal, and leave our conclusions to section V.
  
\section{Topology and CMB}

The topological structure of our universe is an intriguing and fascinating mistery, which has been, and still is, subject of intense theoretical investigations and experimental searches (see~\cite{jl} for a review).  The general questions one goes investigating are whether spacetime has edges, which is the global shape of the universe, what its topological properties.  General Relativity is mostly blind to such questions, for these properties are simply part of the initial conditions of the system.

The simplest, and most often assumed topology physics is constructed in is the usual Minkowski space, with four infinite dimensions spanning $\mathbb{R}^4$.  Cosmology however is somewhat more demanding, and, building on the Cosmological Principles of spatial homogeneity and isotropy, one includes a constant (in space), time-evolving curvature by means of the Friedman-Lema\^itre-Robertson-Walker (FLRW) metric
\be\label{flrw}
\dd s^2 = \dd t^2 - a(t)^2 \left( \frac{\dd r^2}{1 - \kappa r^2} + r^2 \dd \Omega^2 \right) \, ,
\ee
where $\dd \Omega^2 = \dd \theta^2 + \sin^2 \theta \dd \phi^2$, $a(t)$ is the scale factor parametrising expanding physical distances, and $\kappa = 0, -1, +1$ corresponds to spatially flat (euclidean) $\mathbb{E}^3$, hyperbolic (negative curvature) $\mathbb{H}^3$, and spheric (positive curvature) $\mathbb{S}^3$ spaces.  In three dimensions these three options are the only possibilities satisfying the requirements of spatial homogeneity and isotropy, and therefore having constant (in space) curvature.

Within this scheme, the evolution of the universe is dictated by the Friedman equation
\be\label{fried}
H^2 \equiv \left(\frac{\dot a}{a}\right) = \frac{\rho}{3M_P^2} - \frac{\kappa}{a^2} \, ,
\ee
where $H$ is the Hubble parameter, an overdot means a time derivative, and $M_P$ is the reduced Planck mass; the energy density $\rho$ includes any perfect-fluid form of energy density, being it dust, radiation, vacuum energy, or something else. This Friedman equation can be recast as $\kappa = H^2 a^2 (\Omega - 1)$ where $\Omega \equiv \rho / \rho_c$, $\rho_c$ being the critical energy density value for which $\kappa = 0$.  Hence, for instance, if we lived in a universe overfilled with matter, or energy density in general, that is, for $\Omega > 1$, then the global topology of the spacetime would be $\mathbb{S}^3 \times \mathbb{R}$.  Cosmological observations tell us that our spacetime has $\kappa = 0$ to a very good accuracy~\cite{cmb}, so we will employ this value throughout the paper.

In 3d there is a variety of flat (not to be confused with the terminology Ricci-flat!) manifolds that could represent our universe, see again~\cite{jl} for a thorough review on the subject.  In what follows we will focus primarily on the simplest compact manifold, that is, the flat 3-torus $\mathbb{T}^3$, which is obtained by gluing each face of the fundamental cell with its opposite.  Notice that in general one is free to fix the sizes of the fundamental cell, which need not to be equal.  Indeed, a 3-torus is defined by the following identifications
\be\label{torus}
x = x + L_1 \, , \, y = y + L_2 \, , \, z = z + L_3 \, ,
\ee
where the sizes can be ordered as $L_1 \geq L_2 \geq L_3$.  If one or two sides of the torus (in comoving coordinates) are much bigger than the present size of the cosmological horizon than the effective topology would become $\mathbb{T}^2$ or $\mathbb{T}^1$, respectively~\cite{doug,starob}.

Searches for small universes have been performed over the years, mainly exploiting the fact that signatures in the CMB radiation would then be present, in one form or another.  More specifically, direct consequences of a small universe include:
\begin{enumerate}
\item violation of global isotropy as well as global homogeneity;  
\item generically, the spectrum of fluctuations is discrete reflecting the nature harmonics of the finite space; 
\item repeated patterns of temperature fluctuations (hot and cold spots) if some sizes of the fundamental cell is smaller than the size of the last scattering surface (SLS); 
\item a different spectrum of fluctuations which depends directly on the topology through the boundary conditions imposed on the fields living on it (some modes or spherical harmonics may not be available, some other may have a very different shape);
\item a low frequency cutoff in the power spectrum of the temperature fluctuations due to the fact that, in some direction, the lowest multipoles would not fit the fundamental cell.
\end{enumerate}
If we restrict ourselves to the simple case of an effective 1-torus $\mathbb{T}^1$, then it is known that the linear size $L$ of this manifold must be greater than about 24 Gpc~\cite{circles1}, from the non-observation of the patterns dubbed ``circles in the sky''~\cite{circles}.  This limit could be pushed up to around 28 Gps in a the standard cosmological model where vacuum accounts for 3/4ths of the total energy density of the universe, and dark matter makes up the remaining quater of it~\cite{circles2}.  Beyond this value the circles search is not able to return an answer to the question of the finiteness of the universe.  Detecting topology by means of the so called $S$-statistic (see~\cite{oss,teg}) gives similar results, but it is potentially able to go beyond it given better sky maps and improved resolution.

In addition to these limits, the CMB presents a number of anomalies in the low end of the spectrum, which could be pointing towards a non-trivial topology.  Among these are hemispheric asymmetries in the angular power spectrum~\cite{hemi}, a lack of power in the low multipoles (the two-point correlation function for temperature fluctuations seems to vanish for angular separations greater than $60^\mathrm{o}$ circa)~\cite{copi}, and a very special alignment between the quadrupole and the octopole (they both appear to be planar and parallel to each other)~\cite{teg,schw}.  As mentioned, these anomalies could be, and in fact in the literature have been, interpreted as hints to an underlying non-trivial topology, see~\cite{teg,weeks} and references therein.

Notice that most often the analyses on the significance of the above listed anomalies, and their connection to topology, are performed assuming a fiducial gaussian spectrum of primordial perturbations.  Some results may be different if one allows for, e.g., intrinsic non-gaussianity~\cite{pav}.

The new key element which was not available in the previous CMB studies on the topology of the universe~\cite{doug}-\cite{weeks} can be individuated in the new proposed direct relation between the cosmological constant and the topological properties of the manifold: in this scenario \emph{the overall normalisation of topology related effects is unambiguously fixed by the relation~(\ref{rho}), where a minimal manifod size $L$ essentially determines the cosmological constant $\rho_\Lambda$ and vice versa}.  Therefore, eq.~(\ref{rho}) changes the status of CMB searches for a topological structure, which is promoted from ``finding limits on size and shape'' to ``precise measurements'' for this structure.

Let us conclude this section by emphasising once more that \emph{there are no particle physics free parameters in the computations, as everything lies within the standard model}.  In particular, the coefficient $c$ which enters eq.~(\ref{rho}) and which describes how the vacuum energy of a  finite manifold of size $L$ is modified with respect to that of Minkowski flat space, can be computed using standard lattice QCD calculations  as suggested in~\cite{our4d}, for instance by analyzing $1/s$ corrections where $s$ is the total size of the QCD lattice, see the more precise definition below.

\section{The cosmological constant from the ghost}

In what follows we review very briefly the main passages of the mechanism that bridges the well known Veneziano ghost in QCD to the non-vanishing vacuum energy $\rho_\Lambda\neq 0$~\cite{de1,de2}.  This proposal, recently put forward in~\cite{our4d}, is based on the grounds of gravity as an effective field theory, and aims at explaining the late accelerating phase of the universe as a result of a vacuum energy term, incorporated in the Einstein equations in the form of a cosmological constant which, in this scheme, is defined as the mismatch between flat and infinite Minkowski spacetime and a topologically non-trivial one.

If the cosmological constant is indeed a result of the spacetime we live in having a non-trivial topology, then there must be a messenger capable of carrying the information about the boundaries (and the related boundary conditions imposed on the quantum fields), and this is possible only when there are strictly massless degrees of freedom which can propagate at very large distances $\sim H^{-1}$.  The crucial observation is that while na\"ively all QCD degrees of freedom can propagate only to very short distances $\sim \Lqcd^{-1}$, there is a unique (unphysical) degree of freedom which is exactly massless and can propagate to arbitrary large distances: this is the aforementioned Veneziano ghost.  In short, the Veneziano ghost~\cite{veneziano} (see also~\cite{witten}) which solves the $U(1)_A$ problem in QCD is also responsible for a slight difference in energy density between a finite manifold of size $L\simeq H^{-1}$ and Minkowski $\mathbb{R}^4$ space, such that $\rho_{\Lambda}\simeq H\Lqcd^3\simeq (10^{-3} {\text eV})^4$.  Notice that this field is not in the physical spectrum, and as such it does not give rise to all of the usual problems associated with negative sign kinetic terms, propagators, norms and commutation relations.

Thus, the correction to the vacuum energy density due to a very large (but not infinite) manifold is small, and goes as $1/L \simeq H$.  The central point is that although small, this is not exponentially suppressed, $\exp(-L)$, as one could anticipate for any quantum field theory where all physical degrees of freedom are massive (such as in QCD).  Hence, the QCD ghost acts as a source for the cosmological constant, $\rho_{\Lambda}$.  This very small number $(L \Lqcd)^{-1} \simeq H/\Lqcd \simeq 10^{-41}$ nevertheless provides a non-vanishing cosmological constant  which is astonishingly close to the observed value as eq.~(\ref{rho}) explicitly states.

We can not perform an explicit analytical computation including all the parameters describing the exact structure of the manifold in the real world 4d case, not even for the simplest compact manifold such as a torus.  However, the corresponding exact 2d computation can be performed~\cite{toy}.  In that case one finds that the magnitude of the vacuum energy on a large torus of size $L$ slightly changes compared to its Minkowski value as $\sim \frac{\pi}{Lm_{\eta'}} ( \frac{1}{|\tau|} - \frac{1}{\tau_0})$, where $\tau = \tau_1 + i\tau_0$ is the Teichm\"uller parameter for the torus.  This formula exhibits the linear term so intensely sought after $(Lm_{\eta'})^{-1}$.  This result comes from the ghost's contribution, which is very sensitive to the specific boundary conditions at very large distances.  We have all reasons to expect that a similar linear correction $(Lm_{\eta'})^{-1}$ will emerge in 4d QCD analogously to the toy 2d model because of the similarities between their ghost structures. 
 
We now want to give a more precise definition for the coefficient which appears in the expression for the cosmological constant~(\ref{rho}). First of all, there is a dimensionless coefficient $c_{\mathrm{QCD}}$ of order one which is entirely of QCD origin, and is related to the definition of QCD on a specific finite compact manifold such as a torus,
 \be
\label{c_QCD}
\rho_{\Lambda} \simeq c_{QCD}\frac{2 N_f |m_q\la\bar{q}q\ra  |}{L m_{\eta'}} \, . 
\ee
A precise computations of $c_{\mathrm{QCD}}$ can be done in a conventional lattice QCD approach by studying corrections of order $1/s$ to the vacuum energy.  It is obvious that $c_{\mathrm{QCD}}$ depends on the manifold where the theory is defined.  This coefficient is akin to the factor $( \frac{\pi}{|\tau|} - \frac{\pi}{\tau_0})$ which can be analytically computed in the 2d model on a torus with Teichm\"uller parameter
$\tau$~\cite{toy}.

The second factor $c_{grav}$ has a purely gravitational origin and is defined as the relation between the size $L$ of the manifold we live in, and the Hubble constant: $L=(c_{grav} H_0)^{-1}$, such that the final expression for the vacuum energy can be written as
\be
\label{vacuum}
\rho_{\Lambda}  &=& c\cdot\frac{2N_f H_0}{m_{\eta'}}\cdot  |m_q\la\bar{q}q\ra  | \simeq c\,(4.3\cdot 10^{-3} \text{eV})^4 \, ,\\
c&\equiv& c_{\mathrm{QCD}}\cdot c_{grav} \nonumber
\ee
where we use the following standard QCD parameters: $m_{\eta'} \simeq 958$ MeV is the $\eta'$ mass, $m_q \approx 6~\mathrm{MeV}$, and $|\la\bar{q}q\ra| \approx \left( 240~\mathrm{MeV} \right)^3$ is the QCD chiral condensate for two light quarks and $H_0$ stands for the value of the Hubble parameter today.  Let us notice how there is not much QCD-related uncertainty in this expression as the relevant combination $|m_q\la\bar{q}q\ra  |$ can be expressed in terms of the well measured parameters $f_{\pi} \simeq 133$~MeV and $m_{\pi}\simeq 135$~MeV as $4 |m_q\la\bar{q}q\ra  |\simeq m_{\pi}^2f_{\pi}^2$.  Also, we use $N_f=2$ in this expression because the contribution of the $s-$quark into the $\theta$ dependent portion of the vacuum energy is negligible.

The estimate~(\ref{vacuum}) is to be compared with the observational value $\rho_{\Lambda}= (2.3\cdot 10^{-3} \text{eV})^4$ (for $\Omega_\Lambda = 0.73$ and $H_0 = 2.1\cdot10^{-42}h~ \mathrm{GeV}$, $h=0.71$).  The similarity in magnitude between these two values is very encouraging, and may be interpreted as a confirmation that the path we are beating is correct.

A few more comments are in order here.  First of all, the estimate~(\ref{vacuum}) is based on our understanding of the ghost's dynamics: it can be analytically computed in the 2d Schwinger model and hopefully it can be tested in 4d QCD using the lattice QCD computations as explained in~\cite{our4d,toy}.  This contribution to the vacuum energy is computed using QFT techniques in a non-expanding universe.  As it stands, it can not be used for studying its evolution with the expansion of the universe.  In order to do so one needs to know the dynamics of the ghost field coupled to gravity on a finite manifold, that is, solve its equations of motion and compute and renormalise its Green's functions.  Presumably, the dynamics of this field would give us an effective equation of state $w(t)$ which, for this component, is likely to evolve with time.  Presently, we do not know how to implement this dynamics properly within our QFT based framework.

The dynamics of the Veneziano ghost have in fact been discussed from an effective Lagrangian perspective in Minkowski space already by Veneziano himself with Di~Vecchia in~\cite{vendiv}.  In their work, the ghost field is described by means of a non-trivial three-form, which is then combined to form a four-form akin to the topological charge of QCD.  The Lagrangian reads
\be\label{lag}
{\cal L} &=& {\cal L}_0({\cal U}) + \frac{1}{2} i q \Tr\left[ \ln{\cal U} - \ln{\cal U}^\dagger \right] + \frac{N^2}{a f_\pi^2}q^2 - \theta q \nonumber\\
&+& \frac{f_\pi}{2\sqrt2} \Tr\left[ M{\cal U} + M^\dagger{\cal U}^\dagger \right] + g.f. \, ,
\ee
where $g.f.$ means gauge fixing term, ${\cal L}_0({\cal U})$ is the conventional chiral Lagrangian, and the coefficient $a \simeq m_{\eta'}^2$
is fixed by Witten-Veneziano relation for the topological susceptibility in pure gluodynamics. 

In the Lagrangian~(\ref{lag}) the topological density is expressed in terms of a four-form as
\be\label{qdef}
q = \frac{g^2}{64\pi^2} \epsilon_{\mu\nu\rho\sigma} F^{a\mu\nu} F^{a\rho\sigma} \equiv \frac{1}{4} \epsilon_{\mu\nu\rho\sigma} F^{\mu\nu\rho\sigma} \, ,
\ee
with
\be
F_{\mu\nu\rho\sigma} &\equiv& \partial_\mu A_{\nu\rho\sigma} - \partial_\sigma A_{\mu\nu\rho} + \partial_\rho A_{\sigma\mu\nu} 
-\partial_\nu A_{\rho\sigma\mu} \, , \label{four} \\
A_{\nu\rho\sigma} &\equiv& \frac{g^2}{96\pi^2} \left[ A_\nu^a \stackrel{\leftrightarrow}{\partial}_\rho A_\sigma^a - A_\rho^a \stackrel{\leftrightarrow}{\partial}_\nu A_\sigma^a - A_\nu^a \stackrel{\leftrightarrow}{\partial}_\sigma A_\rho^a \right. \nonumber\\
&+& \left.2g C_{abc} A_\nu^a A_\rho^b A_\sigma^c \right] \, . \label{three}
\ee
The fields $A_\mu^a$ are the usual $N^2-1$ gauge potentials for QCD and $C_{abc}$ the $SU(N_c)$ structure constants. 

The three-form $A_{\nu\rho\sigma}$ is an abelian totally antisymmetric gauge field which, under colour gauge transformations with parameter $\Lambda^a$
\be\label{colour}
\delta A_\mu^a = \partial_\mu \Lambda^a + i g C_{abc} \Lambda^b A_\mu^c \, ,
\ee
behaves as
\be\label{transf}
A_{\nu\rho\sigma} &\rar& A_{\nu\rho\sigma} + \partial_\nu \Lambda_{\rho\sigma} - \partial_\rho \Lambda_{\nu\sigma} - \partial_\sigma \Lambda_{\rho\nu} \, , \\
\Lambda_{\rho\sigma} &\propto& A_\rho^a \partial_\sigma \Lambda^a - A_\sigma^a \partial_\rho \Lambda^a \, .
\ee
In this way the four-form $F_{\mu\nu\rho\sigma}$ is a gauge invariant object.  The term proportional to $\theta$ is the usual $\theta$-term of QCD, and indeed the scalar field $q$ is the topological charge density of the system, as anticipated.

Let us point out that the gauge fixing term in~(\ref{lag}) should not be confused with the standard gauge fixing term for the conventional gluon field $A_\mu^a$, since it is related to the fixing of the gauge for the three-form $A_{\nu\rho\sigma}$ describing the Veneziano ghost and carrying no colour indices.  One can understand the field $A_{\nu\rho\sigma}$ as a collective mode which entails a specific combination of the original gluon fields, and which, in the infrared sector of the theory, leads to a pole in an unphysical subspace of the entire Hilbert space.  We know about the existence of this very special degree of freedom and its properties from the resolution of the famous $U(1)_A$ problem, because from integrating out the $q$ field one finds the correct mass of the $\eta'$ meson.

Moreover, not only as far as the $\eta'$ field is under investigation, but in fact in all practical applications the four-form $F_{\mu\nu\rho\sigma}$ is contracted to form precisely by the topological charge density scalar $q$, and then intergrated out by means of its equations of motion, see again~\cite{vendiv} for the details of this procedure.  This is essentially everything one is aiming for from the point of view of the low energy theory in Minkowski spacetime, where no mention to the specificity of the ghost's pole needs to be made, and it is sufficient to calculate all the physical observables of interest.  In order to fully study the dynamics of the ghost in a curved space one would have to identify its proper degrees of freedom in the three-form, and then quantise the theory in the given background.  Unfortunately, especially the second task, appears challengingly complicated and we do not know of any attempt (not to mention successful) in this direction.

Despite this fact, we believe that our result~(\ref{rho}) or~(\ref{vacuum}) is solid and represents a very good approximation of the real world, in that even though we are not able to solve for the dynamics of the three-form, we can see that the ghost's potential is naturally very shallow, due to the linear suppression arising from the compactness of the manifold.  In this case it is very reasonable to expect that the potential energy of the ghost (which is exactly what we have computed) will be of the same order of magnitude of the dynamical term represented by the three-form $A_{\nu\rho\sigma}$.

Hence, to wrap up this long discussion: we have computed the vacuum energy of the ghost in a compact non-expanding universe, which is given by~(\ref{vacuum}).  Since we can not quantise the theory on a non-trivial background there are two uncertainties that arise when we employ this result in the context of our FLRW universe.

The first one is that we do not know what the kinetic contribution is, which, although we expect it to be of the same order of magnitude of the potential term at late times (but still subdominant, and hence not spoiling our estimates for the size of the manifold), it would change the equation of state with time in a way that may not be captured by a simple parametrisation, for instance an approximately constant equation of state, or one with linear dependence on the redshift, $w=w_0+(1-a)w_a$.

The second uncertainty comes from the fact that we can not compute the precise expression for the topological susceptibility in a general curved and compact manifold, for we do not know the expressions for the Green's functions of the ghost in such a setup, due to the intrinsic analytical difficulties of the computation.  Hence, at the moment the equation of state can not be predicted, but the vacuum energy density of the ghost is going to be numerically very close to~(\ref{vacuum}), and we shall retain this formula for our estimates.

The question now is how to explicitly test this idea, which so far appears to be able to explain the tiny value of the cosmological constant, but still has an unknown parameter $c_{grav}$ in it, which can be found only from observations, as it describes the size and topology of the manifold we live in.  Here and in what follows we assume that, although the QCD portion of uncertainty $c_{\mathrm{QCD}}$ presently can not be computed analytically, still is not treated as an unknown coefficient.  Indeed, QCD is part of the standard model and consequently all parameters are in principle calculable from first principles by using standard techniques, e.g.\ numerical lattice computations.  So, to simplify our notations in the remaining part of the text we set $c_{\mathrm{QCD}}=1$.  Once the QCD calculation for a given manifold is performed, the modifications to our results are easily implemented.

It turns out that the CMB may very soon be able to test this proposal and either gather data in support of it or reject it, as is explained in the next section.

\section{Observing topology in the CMB}

The most immediate consequence of expression~(\ref{vacuum}) is that if the cosmological constant $\rho_{\Lambda}$ indeed arises from the finiteness of the manifold we live in, than the corresponding topological structure on the scale $1/L\simeq H_0$ can be probed using the last scattering surface (SLS) imprinted in the CMB.  Therefore, since the dark energy and the topological structure of the universe are intimately linked one another, it is possible to try to measure one of them in order to obtain information on the other one.  In particular, we would be looking for a non-trivial topology in the microwave sky, whose typical size is set by its relation to the observed vacuum energy.

The only residual information coming from the fact that the manifold we live in could be compact is stored in the free constant $c$ of formula~(\ref{vacuum}).  This constant does not specify exactly which manifold we are dealing with, because, as previously mentioned, it is not possible to track analytically the parameters defining it.  However, once more referring to the 2d model of~\cite{toy}, this coefficient is expected to be entirely specified by the structure of the manifold, such as its linear sizes, the angles at which the sides are with respect to each other, and the relative twisting of the glued faces.  In particular, it arises whenever there is an asymmetry between different linear lengths, in which case the linear correction appears.

Hence, the precise structure of the manifold is not fully assessable; nevertheless, we can estimate its linear size by comparing, or normalising, the expected mismatch in vacuum energy eq.~(\ref{vacuum}) to the observed one.  This linear size would then refer to the smallest dimension, and would describe an effective $\mathbb{T}^1$-universe.  Practically, one can define this size of the manifold as $L = (c_{grav} H_0)^{-1}$, and therefore explicitely obtain an estimate for the linear length of the torus
\be\label{size}
L = \frac{1}{c_{grav} H_0} \approx 17 H_0^{-1} \approx 74 \mathrm{Gpc} \, .
\ee
Notice that this number, although subject to some variability when the reference numbers are chosen slightly differently, is beyond that which can be probed with the circles in the sky method.  Moreover, the current CMB mission Planck is most likely going to be able to look for signatures of a small universe of such size, thanks to improved resolution over that of COBE and WMAP available for the analyses~\cite{oss,teg}.

A different way of putting the result quoted in~(\ref{size}) is by comparing it to the size of the SLS, which, in a FLRW universe filled with dust and cosmological constant in respective proportions of $\Omega_M = 0.27$ and $\Omega_V = 0.73$, is written as
\be\label{sls}
d_{\mathrm{SLS}} = a_0 \! \int_{t_{\mathrm{SLS}}}^{t_0} \!\frac{\dd t}{a(t)} \simeq \frac{2}{\sqrt{\Omega^0_M} H_0} {}_2F_1(\frac{1}{6},\frac{1}{2},\frac{7}{6},-\frac{\Omega^0_V}{\Omega^0_M}) \, .
\ee
With this definition we find
\be\label{size2}
d_{\mathrm{SLS}} \approx 3.4 H_0^{-1} \Rar L \approx 5 d_{\mathrm{SLS}} \, ,
\ee
which is well beyond the entire SLS given by $2d_{\mathrm{SLS}}$.

Another consequence of this proposal is its sensitivity to the orientation of the manifold.  One can measure the so called integrated Sachs-Wolfe effect (ISW) when the gravitational potential varies with time between the SLS and now.  In this case any compact manifold with a specific orientation gives rise to the ISW effect.  If our proposal is correct, we predict that for a given linear length as estimated in~(\ref{size}) there should be a specific orientation for which the correlation pattern describes the data much better than the standard description in Minkowski flat space. In fact, the relevant technique has been developed~\cite{BondA,BondB} for a torus and other simple manifolds, and we encourage new analyses taking into account the fact that the parameter~(\ref{size}) is now fixed (if the model is correct).
 
We claim that such kind of asymmetry due to the ISW effect (if observed) is closely related to another type of asymmetry of the CMB which would also arise in our model, and which apparently has been already observed, see the recent paper~\cite{hemi} and the extensive list of  references on previous works therein.  The main result of ref.~\cite{hemi} is that the hemispherical power asymmetry previously reported for the largest scales $l = (2-40)$, where $l$ is the $l$-th multipole in the standard harmonic decompisition of the CMB spectrum of temperature fluctuations, extends to much smaller scales.  In fact, for the full multipoles range $l = (2-600)$ significantly more power is found in the hemisphere centered at ($\theta = 107^o\pm 10^o, ~ \phi = 226^o\pm 10^o$) in galactic co-latitude and longitude, than it is in the opposite hemisphere.  We remark that this specific orientation must be correlated with that extracted from the analysis of ISW effect as discussed above, because in our model the source of both asymmetries is one and the same (the orientation of a compact manifold).
   
One more consequence of a finite universe is the presence of a low frequency cutoff in the power spectrum of the temperature fluctuations due to the fact that, in some direction, the lowest multipoles would not fit the fundamental cell.  Observationally it implies that the multipoles with lowest $l$ (such as the quadrupole) will be affected the most as they are the most sensitive to the largest scales.  As mentioned, the lack of power in the quadrupole could be interpreted as a hint to an underlying non-trivial topology, see~\cite{teg,weeks} and references therein.

Finally, the estimates~(\ref{vacuum}) and~(\ref{size}) give a very specific prediction for the level at which Lorentz invariance will be broken, as any compact manifold size $L$ obviously breaks the Lorentz symmetry, see e.g.\ the review paper~\cite{Klinkhamer:2005et}.  Notice that this violation of Lorentz symmetry will appear in the metric at the level of $1/L$, which in turn means that the energy momentum tensor, in our case proportional to the metric tensor, will not necessarily have the perfect fluid, cosmological constant form.  This is seen explicitely in the 2d toy model discussed in~\cite{toy}.

It is important to stress once more that thanks to the proposed mechanism that explains the observed value for the vacuum energy we are able, via its link with the possible non-trivial topology of the 3d space, to make a prediction, eq.~(\ref{size}) or~(\ref{size2}) on the size of the compact manifold.  This number is entirely determined by known QCD physics.   If the value of the cosmological constant is set by this mechanism, then the size of the manifold is also fixed, rendering the mechanism directly testable and falsifiable through CMB measurements.

\section{Conclusion}

In this letter we have discussed the most immediate consequence of the recent proposal of~\cite{our4d}, that is the existence of at least one compact dimension within our 4d spacetime.  The mechanism described in~\cite{our4d} provides a tentative explanation for the cosmological constant puzzle based solely on QCD arguments, more specifically, reanalysing the solution of the axial $U(1)_A$ problem as formulated by Veneziano, exploiting the properties of the so-called Veneziano ghost field in a general compact manifold.

The observed vacuum energy is described, within this framework, as a mismatch between that obtained in a fully infinite Minkowski space and a topologically non-trivial manifold such as a 3-torus.  In this case the value of the vacuum energy is almost entirely written in terms of well known QCD parameters, the only exception being the size of the compact manifold the model demands we live in.  However, knowing the numerical value of the observed dark energy, one is in a position to fix this unknown parameter univocally.  If this mechanism is indeed the one responsible for the appearance of the cosmological vacuum energy then the linear size of the manifold is predicted to be about 74 Gpc, or about 5 times the SLS radius.

The exact structure of the manifold is not determined by the arguments given in~\cite{our4d}, and the corresponding fine signatures in the CMB are therefore not given.  Moreover, our field theoretical calculations do not include the expansion of the universe, and therefore do not describe entirely accurately our world.  However, it is still possible to look for such a ``small'' universe in microwave maps in quite a general way by studying the correlated low frequency cutoffs and power spectra.  Indeed, data in favour of some sort of preferred direction and missing long wavelength power has been accumulating since COBE, and confirmed by WMAP.  PLANCK will be able to confirm or falsify the results obtained in this work.

\section*{Acknowledgements}
We thank P.~Naselsky for discussions on the observed asymmetry in the CMB, D.~Scott for valuable conversations, A.~Starobinsky for explaining his model~\cite{starob} and F.~Klinkhamer and Grisha Volovik for correspondence.  This research was supported in part by the Natural Sciences and Engineering Research Council of Canada.

\end{document}